# The Structure of Information

Bruce Long, Dept. of Computer Science, University of Westminster, London
B.D.Long@westminster.ac.uk
September 1, 2003 (Draft)

## Abstract

A formal model of the structure of information is presented in five axioms which define identity, containment, and joins of infons. Joins are shown to be commutative, associative, provide inverses of infons, and, potentially, have many identity elements, two of which are multiplicative and additive. Those two types of join are distributive. The other identity elements are for operators on entwined states. Multiplicative joins correspond to adding or removing new bits to a system while additive joins correspond to a change of state. The order or size of an infon is defined. This groundwork is intended to be used to model continuous and discreet information structures through time, especially in closed systems.

## Introduction

In this paper we present a formal model of information. The object of this model is intended to be evolving physical state systems reinterpreted via the equation converting states (s) to bits (b), $s = 2^b$. That is, every state system, whether classical or quantum, will be considered to be information. There are a number of advantages to such a scheme. With an evolving state model of a system confusion can arise about the identity of the system; under a philosophically naive view there is a single system that somehow "changes" over time. A more formal view usually represents a series of "snapshots" of the system separated by "events." That is, there are many different systems, each slightly different from the previous one, and each of which does not change itself. However, consider that in a deterministic system, the information in the system does not change. Its identity remains over time. For example, while a CD player may "change" state to emit the sounds of an orchestra, the information in the airwaves is identical to (or contains a component identical to) the information from the orchestra entering the recording system. Likewise, when an algorithm is run, or when a quantum system evolves, no new information is added to the system during the run or experiment (this is the meaning of "deterministic system"); rather, the information is structured in a certain way that observers, as part of that structure, perceive as the flow of time. Each piece of information in a deterministic system exists at every "instant" of the structure. In addition to modeling more about the structure of evolution than a typical state-based model, the formalism presented here models continuous and closed information systems, and thus may be of interest to quantum physicists as well as computer scientists and mathematicians. For example, while a non-closed eight-bit system (a byte) might not evolve at all, or might evolve to successively take on all the values of bytes in an entire database, a truly closed eight-bit system can only take on 256 distinct configurations. One such configuration would be to not evolve at all. Other such configurations would allow evolution, but at each instant fewer than eight bits would be expressed. We shall see that while there are only 256 distinct configurations, there are many more ways other systems can interact or observe that system.

The companion paper to this one, "Quanta, a Language for Modeling and Manipulating Information Structures", builds upon the theory presented here such that macro systems can be formally represented, queried and manipulated. The current formalism is best for modeling very small, closed systems. Goals for the theory are that most of the common properties of information, such as those in Shannon's information theory -- which deals with the *amount* of information, not its *structure* or *content* -- arise from the theory without being built in. For example, the fact that states and events exist, and the relation between the number of states and the number of bits of information should arise from the theory. The two major assumptions made in the current theory are that information can be divided into sub-parts, for example, a byte can be divided into two four bit chunks, and that any two pieces of information are either identical to each other or are not identical to each other.

## Overview

We borrow the term "infon" from Keith Devlin as was coined in his *Logic and Information* [Devlin], to mean "piece of information." We could define infons formally to be whatever is described by the theory, or we could refrain from defining them formally and rely on the reader's knowledge of what information is. We choose the latter definition because our intention is to describe actual occurrences of information, and as such, the theory should be falsifiable. Whether the reader's view of information is the computer science view of files, memory locations and data structures, or the physics view of evolving state systems, or the mathematics view of abstract entities, if the following theory does not describe the phenomenon, then, unlike with pure theory, we will consider the theory to be flawed and search for a problem.

We start by describing infons in much the same way that sets are described in set theory. There are five axioms: two defining identity in terms of substitutability, one defining containment, one for sub-members, and one defining joins of infons. We define intersection of infons and infer that, unlike sets, infons may sometimes be non-identical yet have the same members. For example, a symmetrical shape interacted with from a different position contains the same parts but the interaction is still different in an important way. From these axioms we infer that infons are associative, commutative, and that they have may have many identity members depending upon the size of their intersection. Infons have inverses around the identity members. Next we show that when the intersection between two infons is null, the infons behave multiplicatively such that two four-state infons join to become a 16-state infon. Likewise, when the infons fully intersect their joins behave additively. Thus, '1' and '0' infons are defined as identity members for multiplicative and additive joins. The other identity members, which define entwined states, will be discussed in a future paper.

# Formalism

**Infons**

*Def: Infons* are pieces of information.

**The Identity Relation**

For any two infons A and B, we denote an *identity* relation with: A==B. We may also say "A is B" or "A and B are the same infon."

We denote non-identity with: A!=B.

*Axiom of Identity and Substitutability*: For any two infons A and B, if A==B, then truth (or information content) will be preserved if, in an extensional context, A is physically replaced with B.

*Axiom of Excluded Middle*: For any two infons A and B, either A==B or A!=B, but never both.

*Theorems of Identity*: Identity is reflective, symmetric and transitive.

> Proofs: Suppose that there exists an A such that A!=A, and let there be an infon B such that A==B. Then by substitution we get A!=B; but by *excluded middle* this cannot be, so A==A (reflectivity). Let A and B be any two infons where A==B. Then substituting B with A and A with B we get B==A (symmetry). Let A, B and C be any two infons such that A==B and B==C. Then by substitution of B with C we get A==C (transitivity).

Note: Equality does not imply identity. For example, if A==5 where A is the number of rooms in a house, and B==5 where B a person's age, then while A=B, A!=B. A=B is compatible with either A==B or A!=B.

**The Containment Relation**

For any two infons A and B, we denote that A *contains* B with: A : B. We may also say that B is a member of A or that B is a sub-infon of A.

*Axiom of Containment and Identity*: If any two infons A and B have the same members, then A==B. (Where sameness denotes identity, not equality).

*Theorem of Identity and Containment*: For any two infons A and B, if A==B, then A and B have the same members.
Proof: Let A, B and D be any three infons where A : D and A==B. Then substituting A with B, we get B : D. Likewise for B : D where B is substituted with A.

*Axiom of Recursive Containment*: For any two infons A and B, A contains B iff every member of B is also a member of A.

*Terminology*: To simplify terminology, if we say "A contains B and nothing else," we mean that A contains B *and* every infon implied by recursive containment and the other axioms, but nothing else.

Note: This axiom makes containment a meriological relation, e.g., a book contains all of its pages, all the parts of those pages, all their cells, molecules, atoms, etc. This does not usually complicate proof because, for example, if an infon contains A and B and all their descendents, one only need prove that A and B are contained to prove identity; containment of the descendents automatically follows.

**Null Infons**

Def: If all the members of an infon A are identical to each other, A is called a *null infon*.

Notes: Readers may question why a "null infon" contains a single member, itself, rather than no members. Because members are analogous to states in a system, a null infon contains only one state, which is, of course, zero bits of information. The name "null" is chosen to signify that null infons are void of any information. Null infons play a role in this theory which is similar to the role that null *sets* play in set theory.

**Intersection of Infons**

*Def: Intersection*: For any two infons A and B, an infon I(AB) refers to the infon which contains all, and only, members of A that are identical to members of B.

*Def: Disjoint Infons*: For any two infons A and B, A and B are *disjoint* if they have no members other than null in common, that is, if I(AB)==null.

*Def: Covering Infons*: For any two infons A and B, A and B *cover* each other iff every member of A is identical to a member in B and every member of B is identical to a member in A. (Because containment of the same members is not sufficient for inferring identity, the concept of a cover is needed.)

**Joining Infons**

*Axiom of Joining*: For any two infons A and B, an infon AB (called the *join of A and B*) exists which contains A and contains B, but nothing else. (See note under the *axiom of recursive containment*.)

*Theorem of Commutativity of Joining*: For any two infons A and B, AB==BA.

> PROOF: For any two infons A and B, AB contains A and B, and nothing else; likewise, BA also contains A, B, and nothing else (axiom of joining). Since they contain the same members, AB==BA (axiom of containment and identity).

*Theorem of Associativity of Joining*: For any three infons A, B, and C, (AB)C == A(BC).

> PROOF: For any three infons A, B and C, AB contains A and B, and nothing else; joining that to C adds C and nothing else, giving A, B, and C as members. Likewise, BC contains B, C and nothing else. Joining A to that adds A and nothing else, giving A, B, and C as members (axiom of joining). Since they contain the same members, (AB)C == A(BC) (axiom of containment and identity).

Note: The previous two theorems could be special cases of a theorem that joining does not add information; for example, suppose A and B are disjoint and are each four-bit infons. If information is conserved in a join, then for AB==D, D should be eight-bits, but if the ordering of the infons were stored in D, then D must be at least 9 bits. Thus, since information is conserved in a join, the order of the infons must not be stored in D.

*Theorem of One-To-One Joins*: For any infon C and for any three of its members A, x and y, if I(Ax)==I(Ay) then [Ax==Ay iff x==y].

> Proof: Suppose x != y. Then there exists an infon that is a member in either x or y but not both (*containment and identity*). Suppose x is the infon with the extra member. There are two possibilities: either the extra element is also in A and therefore in Ay, or it is not in Ay. If I(Ax)==I(Ay) then the element is not in A because we supposed that it is not in y, yet the intersection between A and y is also in A, so the extra element is not in Ay (*recursive containment*). Thus when I(Ax)==I(Ay) and x!=y, Ax != Ay, i.e., A maps x 1-to-1 to Ax (*containment and identity*).

*Theorem of Onto Joins*: (By keeping A constant, but allowing x to vary, Ax==y can be viewed as a function A(x)=y mapping x to y.) Let D be any infon and let A and B be members of D. If the domain and codomain of A() are all w in D such that I(Aw)==I(AB) then *Ax==y* maps x *onto* y.

> Proof: Since the domain and codomain of A(x) are the same and are such that their intersections with A are all identical, and because if I(Ax)==I(Ay), A maps x 1-to-1 to y (*One-To-One Joins*), the range of A includes every member of the codomain which is the definition of "onto."

*The Sibling Theorem* (joining does not remove information): For any infon G, and for any two members of G, A and C, there exists a member of G, B, such that AB==C.

> Proof: (By keeping A constant, but allowing x to vary over the members of G, Ax can be viewed as a function A(x)=y mapping x 1-to-1 onto y (*One-To-One Joins, Onto Joins*).) This means that for every member of G, there is at least one value for x which maps A to that member. Since G is a member of itself (recursive containment), the Sibling Theorem is true.

**Identity Elements and Inverses**

We look at two special cases of the Sibling Theorem. First we apply the Sibling Theorem to the case of AB==C where A==C. (Note that there exists a member of C, A where A==C because any infon is a member of itself.)

*Def: Identities for Single Infons*: For any infon G, and for any member of G, A, we refer to a member of G, *i*, that exists (by the sibling theorem) in the expression A*i*==A as an *identity member* of A in G.

Second, since an identity member for any given infon exists, we consider the expression AB==*i* for the member A in some infon G.

*Def: Inverses*: For any infon G, and for any member of G, A, we refer to the member of G, B, that exists in the expression AB == C (where C is the identity of A in G) as the *inverse of A in G*, or, as long as it is known that G is the relevant parent for A, $A^{-1}$. Thus, if A is a member of G, A $A^{-1}$==*i*.

*Theorem of Identity Groups*: For any infon G, and for any two members of G, A and B, let *i* and *j* be the identities of A and B respectively. I(A *i*)==I(B *j*) iff *i*==*j*.

Proof: Let A and B be members of any infon G, and let *i* and *j* be the identities of A and B respectively. We start with (A *i*)B==A(*i* B). Since A *i*==A, we have AB==A(*i* B). Viewing A as a function we have A(B)==A(*i* B). Since, when I(A *i*)==I(B *j*), A is a 1-to-1 function that maps onto its codomain, B==(*i*B)==(B*i*). Thus, *i*==*j*. If *i* != *j*, then A() is not acting as a 1-1 onto function so it must be that I(A *i*) != I(B *j*).

**Identity Groups**

Thus far we have shown that the join operation for infons which have a particular intersection is much like a group operation. However, as the intersection of the operands in a join varies, so does the member of the infon which acts as the identity member for the join. We look at the cases for both disjoint infons and for covering infons.

*Def: Multiplicative Joins*: For any two infons A and B, if A and B are disjoint, we refer to their join as *A * B*. We refer to the identity member for disjoint joins with the symbol "1". We may refer to the join of A with the inverse of B by *A/B*. Thus, the multiplicative (disjoint) inverse of an infon A can be referred to as *1/A*.

*Def: Additive Joins*: For any two infons A and B, if A and B are covers, we refer to their join as *A+ B*. We refer to the identity member for disjoint joins with the symbol "0". We may refer to the join of A with the inverse of B by *A-B*. Thus, the additive (covering) inverse of an infon A can be referred to as *0-A* or *-A*.

Note: We will not discuss the other identity members here. The joins of infons with varying degrees of intersection can be decomposed into combinations of additive and multiplicative joins.

The notations A*B and A+B assert not only a join but identity or non-identity. Because of this double assertion, these operations are not associative when combined with each other.

*Theorem of Distribution*: For any three infons A, B, and C, A*(B+C) == (A*B) + (A*C)

   Proof: *Not available in this draft.*

*Def: Literal Representation*: Though there are important differences between infons and numbers, the description of infons is sufficiently similar to numbers that we shall use the decimal numbering system to represent infons and their membership.

   Example: An infon 12 contains disjoint members 2*6, 3*4, etc. It contains covered infons 5+7, 2+10, 5.5+6.5, etc. Note, that identity must be tracked separate from equality. For example, the infon A==35 may record one's age while another infon D==35 may record the distance one must travel to work. While they are equal, they are not identical.

*Def: Order Relations*: For an infon G, we define the order relations < and > in the usual way.

**Non-Identities and the Size of Infons in States**

In addition to specifying identities, an important aspect of describing infonic structure is specifying non-identities. For example, any structure of identities is true of the null infon since all of its members are identical.

The covers of an infon B are all those infons which are the result of addition joins with B. Since any two infons are either identical or not identical to each other, the question arises as to, what identities and non-identities are possible among the covers of an infon.

*Theorem of Finite Infons*: For an infon G, for any cover of G, E, if E==0 then by substitution, for any cover of G, F, *E+F==0+F*.

*Def: Order of an Infon*: For an infon A, the lowest member of A, B>=1 such that B==0 is called the *order* of A, and denoted by A_o where o is the order. For example, 5_8 is an order 8 infon with value 5.

Note that iff the order of an infon is 1, that is, 0==1, then all its members must be identical and thus the infon is null. The order of an infon corresponds to the number of states in that infon.

*Theorem of Order of Joins*: For any two infons A_n and B_m, the order of AB is mn.

   Proof: *Not available in this draft.*

For example, two disjoint infons 3_4 and 4_5 join to 12_20, an infon in state 12 of 20. When two covering infons are joined, their initial orders are the same (since they are covers, they are the same size). Under addition, their order is also the additive identity element, 0. Thus, the result order is also the same. E.g., 1_4 + 2_4 == 3_4.

*Theorem of Non-Identity*: For any infon G, for any two integral covers of G, A and B such that 0<=A<A_O and 0<=B<B_O and A<>B, A!=B.

## Future Research: Representing Temporal Structures

It is intended that future research will look at how the linear structure of infons generates temporal structures and how the continuous nature of infons (due to the fractional members between integral states) produces information in the form of waves. We will look at how to represent sequences, repetitions and conditionals.